\documentclass[letter]{aa} 

\usepackage{graphicx}
\usepackage{natbib}
\usepackage{txfonts}
\usepackage[colorlinks=true,citecolor=blue]{hyperref}
\usepackage{xcolor}
\begin{document}

\title{Angular momentum transport near convective-core boundaries of Gamma Doradus stars}

      \author{F.D. Moyano
        \and
        P. Eggenberger
        \and
        S.J.A.J. Salmon
      }
          
      \institute{Observatoire de Gen\`eve, Universit\'e de Gen\`eve, 51 Ch. Pegasi, CH-1290 Versoix, Suisse
        \\ email: facundo.moyano@unige.ch
      }
          \date{Received: 22 November 2023; accepted: 25 December 2023}
          \titlerunning{On the inner differential rotation of main sequence stars}
 
          \abstract
              {Recent asteroseismic studies have revealed that the convective core of $\gamma$ Doradus stars rotates faster than their radiative interior.
              We study the development of differential rotation near the convective core to test angular momentum transport processes that are typically adopted in stellar evolution models.
              Models that only include the advection of angular momentum by meridional circulation and shear instabilities cannot reproduce current rotational constraints, irrespective of the initial conditions.
              The latest formulation of internal magnetic fields based on the Tayler instability is indeed able to reproduce the internal rotation rate of post-main sequence stars, however, it appears too efficient during the main sequence and has thus been disfavoured.
            A less efficient version of the same transport process can  simultaneously reproduce the rotation rate of the convective core, the rotation rate in radiative regions as probed by gravity-modes, and the surface rotational velocities of $\gamma$ Doradus stars.
            Our work suggests that there are additional physical processes apart from internal magnetic fields at work in the stellar interiors of post-main sequence stars.
            }


   \maketitle
%
   \section{Introduction}
   \label{intro}
   A number of recent results point to a misunderstanding with respect to the modelling of rotation in stellar interiors through evolution \citep[e.g.][and references therein]{aerts19}.
   Since the theory of angular momentum (AM) transport by meridional circulation and shear instabilities has failed to account for observational constraints brought by asteroseismology \citep[e.g.][]{eggenberger12}, this has led to the conclusion that additional physical processes would transport AM efficiently in stellar interiors.
   Currently, the theory of internal magnetic fields generated by the Tayler-Spruit dynamo \citep[hereafter, TS;][]{spruit02} in its revised form \citep{fuller19} is the closest and most widely tested solution to the AM transport problem in stellar evolution theory.
   However, disagreements with regard to the core rotation rates of subgiants have undermined this solution \citep[e.g.][]{eggenberger19}.
   The recent detection of radially differential rotation between the convective core and the radiative interior of main sequence (MS) stars \citep{saio21} poses an additional, yet fundamental, challenge to our understanding of AM transport in stellar interiors, since asteroseismic studies of MS gravity-mode pulsators had provided the rotation rate only in their radiative regions.

   In MS stars, particularly Gamma Doradus stars (hereafter $\gamma$ Dor), an almost uniform rotation is favoured along the radiative interior \citep[e.g.][]{vanreeth18}, although differential rotation in the near-core regions could not be excluded.
   The rotation rate of the convective core was recently measured for the first time in a sample of 16 $\gamma$ Dor stars \citep{saio21} by modelling features called `dips' seen in the period spacing patterns of gravity modes \footnote{A period spacing pattern is the relation between the difference in period, $\Delta P_{\rm n}=P_{\rm n+1}-P_{\rm n}$, between modes of consecutive radial order and the pulsation period, $P_{\rm n}$, of different radial modes.}.
   As shown by \citet{ouazzani20} those dips are caused by resonant coupling between inertial modes trapped in the convective core and gravity modes propagating in the radiative interior \citep [See also][]{tokuno22,aerts23}.
   What is crucial to glean from these first constraints is that the convective core rotates slightly faster than the radiative interior, with a difference of at most $\sim 20 \%$.
   Although this may appear to be a small difference, in combination with previous results, it provides strong information on how the physical processes should operate in stellar interiors. 
   Moreover, these constraints allow us to study the AM transport in regions with strong chemical gradients, since the mass of the convective core in these stars is expected to decrease through the MS, thereby  leading to a gradual change in the chemical composition between the convective core and the radiative interior.
   This makes it particularly important to carry out detailed tests of the inherent physical processes, since several hydrodynamic and magnetic instabilities in stellar interiors are inhibited by chemical composition gradients. This is because the fluid motions are expected to be able to counteract the buoyancy forces \citep[e.g.][]{heger00}. 
     
   In this paper, we study the development of differential rotation between the convective core and the radiative interior through the MS in models with purely hydrodynamical processes, namely, the advection of angular momentum by meridional currents and shear instabilities, as well as in models that also include internal magnetic fields generated by the TS dynamo.
   We further test whether these models are in agreement with the estimates provided by \citet{saio21} and we consider their applicability to more advanced phases.
   
   \section{Input physics of models}
   \label{physics}
   We used the Geneva stellar evolution code \citep[{\fontfamily{qcr}\selectfont GENEC};][]{eggenberger08} to compute models of rotating stars during the MS.
   The code includes a detailed treatment of AM transport following the formalism of \citet{zahn92}, which takes into account AM advection by meridional currents and diffusion by turbulent processes such as shear instabilities.
   We employed the formulation of \citet{maeder97} for the secular shear instability and the formulation presented by \citet{eggenberger22} for the internal magnetic fields generated by the Tayler instability.
   We assumed that convective overshooting mixes chemical elements instantaneously, but the temperature gradient remains non-adiabatic in the overshooting regions (i.e. the step-overshooting formalism).
   We used an overshooting strength of $d=0.05 H_{\rm p}$ (where $H_{\rm p}$ is the local pressure scale height) for models with $M < 1.7 M_{\odot}$, and  $d=0.1 H_{\rm p}$ for models with $M \ge 1.7 M_{\odot}$.
   The rest of the input parameters and the physics included are the same as those given in \citet{moyano23b}, unless otherwise specified, and we refer to this work for further details.

   The AM transport by internal magnetic fields in our models relies on a generalised prescription of the Tayler instability as given by \citet{eggenberger22}.
In this prescription, the uncertainties on the timescale needed to damp the instability on the azimuthal magnetic field is parameterised through a variable $C_{\rm T}$.
The general expressions that determine the efficiency of AM transport are hereafter given by Eqs. \ref{eq_qmin} and \ref{eq_numag}.
   The former determines the condition needed for the dynamo to operate which requires a minimum degree of local differential rotation, termed $q_{\rm min} \equiv |\partial \log \Omega / \partial \log r|_{\rm min}$, and given by:
   \begin{equation}
     \label{eq_qmin}
     q_{\rm min}=\frac{1}{C_{\rm T}} \left ( \frac{N_{\rm eff}}{\Omega} \right )^{(n+2)/2} \left ( \frac{\eta}{r^2 \Omega} \right )^{n/4}
   ,\end{equation}
   where $N_{\rm eff}$ is the effective Brunt-V\"ais\"al\"a frequency, $\eta$ is the magnetic diffusivity, $\Omega$ is the angular velocity, $r$ is the radial distance from the centre of the star, and $n$ is a parameter that determines the damping timescale of the instability.
   And the latter sets the efficiency of the AM transport which is regulated by the magnetic viscosity ($\nu_{\rm mag}$) and is computed as
   \begin{equation}
     \label{eq_numag}
     \nu_{\rm mag} = \frac{\Omega r^2}{q} \left ( C_{\rm T} q \frac{\Omega}{N_{\rm eff}} \right )^{3/n} \left ( \frac{\Omega}{N_{\rm eff}} \right ).
   \end{equation}
   We recall that with $C_{\rm T}=1$ and $n=1$ the formalism reduces to that originally proposed by \citet{spruit02} and $n=3$ to the one proposed by \citet{fuller19}.
   In this work, we use $n=1$ and different values of $C_{\rm T}$ to study the degree of differential rotation between the convective core, and the near-core rotation rate in the gravity-mode cavity probed with asteroseismic techniques, since it sets the efficiency of AM transport; a lower(higher) value of $C_{\rm T}$ leads to less(more) efficient AM transport.

   We compute the near-core rotation rate ($\Omega_{\rm nc}/2\pi$) as an average rotation rate of the gravity-mode (g-mode, hereafter) cavity from our models as
  \begin{equation}
    \label{eq_meanomega}
    \frac{\Omega_{\rm nc}}{2\pi}= \frac{1}{2\pi} \frac{\int_{r_{\rm in}}^{r_{\rm out}}{\Omega N dr/r}}{\int_{r_{\rm in}}^{r_{\rm out}}{N dr/r}}
  ,\end{equation}
  where $N$ is the Brunt-V\"ais\"al\"a frequency, and $r_{\rm in}$ and $r_{\rm out}$ are the inner and outer limits of the g-mode cavity, respectively.
  The rotation rate of the convective core ($\Omega_{\rm cc}/2\pi$) is simply taken as the rotation rate of the centre of the model since we assume rigid rotation in convective regions.
  In the following sections, all the models presented have an initial mass of $M=1.5 M_{\odot}$, an initial metallicity of $Z=0.01$, and an initial rotation rate of $\Omega/2\pi = 20\mu$Hz at the zero-age main sequence (ZAMS) assuming initial solid body rotation, unless otherwise specified.

     \begin{figure}[ht!]
     \resizebox{\hsize}{!}{\includegraphics{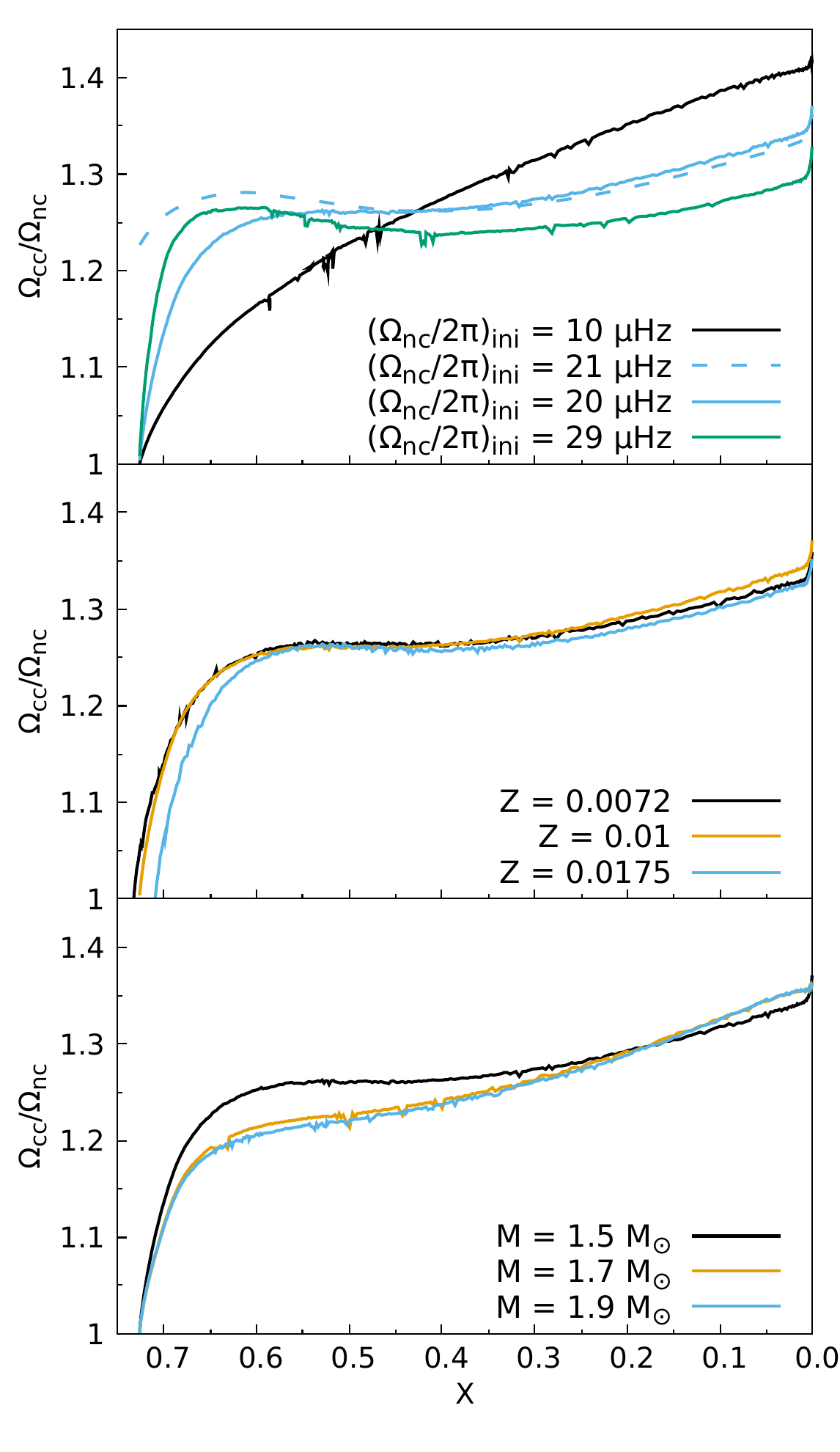}}
     \caption{
       Ratio of rotation rate between the convective core and the near-core radiative interior probed by gravity modes, as a function of the central hydrogen abundance in mass fraction for models without internal magnetic fields.
       The effect of initial rotation rates (top), metallicity (middle), and initial mass (bottom) are illustrated in each panel.
       All models are computed assuming uniform rotation as an initial condition, except for the model in light-blue dashed line in the top panel where we assumed that initially the convective core rotates faster than the near-core regions.
     }
     \label{innerdr_xc_multiplot}
     \end{figure}
     %
     %
     \section{Models with no internal magnetic fields develop too much differential rotation}
  Models without internal magnetic fields always develop differential rotation between the convective core and the near-core radiative interior (hereafter inner differential rotation) early during the MS.
  In Fig. \ref{innerdr_xc_multiplot} we show the degree of inner differential rotation as measured by the ratio of the rotation rate between the convective core and the near-core radiative region ($\Omega_{\rm cc}/\Omega_{\rm nc}$) during the MS.
  The initial rotation rate (and hence the rotational velocity) of the models affects the degree of inner differential rotation more strongly than the initial mass or the metallicity.
  While metallicity has a negligible effect, the initial mass has a mild effect, leading to a difference of at most 5\% in the inner differential rotation during the early MS. 
  Low initial velocities lead to higher degree of inner differential rotation during the late MS (central hydrogen mass fraction of $X \lesssim 0.35$), whereas high initial velocities lead to a rapid decoupling between the convective core and the radiative regions as a result of the higher efficiency of advection by meridional currents at higher rotation rates.
  We also verify whether it is possible to progressively decrease the degree of inner differential rotation by starting with $\Omega_{\rm cc}/\Omega_{\rm nc} \simeq 1.25$ at the ZAMS (shown by the light-blue dashed line in Fig. \ref{innerdr_xc_multiplot}). This behaviour was suggested by \citet{saio21} due to the correlation that these authors found between the inner differential rotation and the central hydrogen abundance.
     We find that this scenario is not achievable in our models with purely hydrodynamical processes.

\begin{figure}[htb!]
     \resizebox{\hsize}{!}{\includegraphics{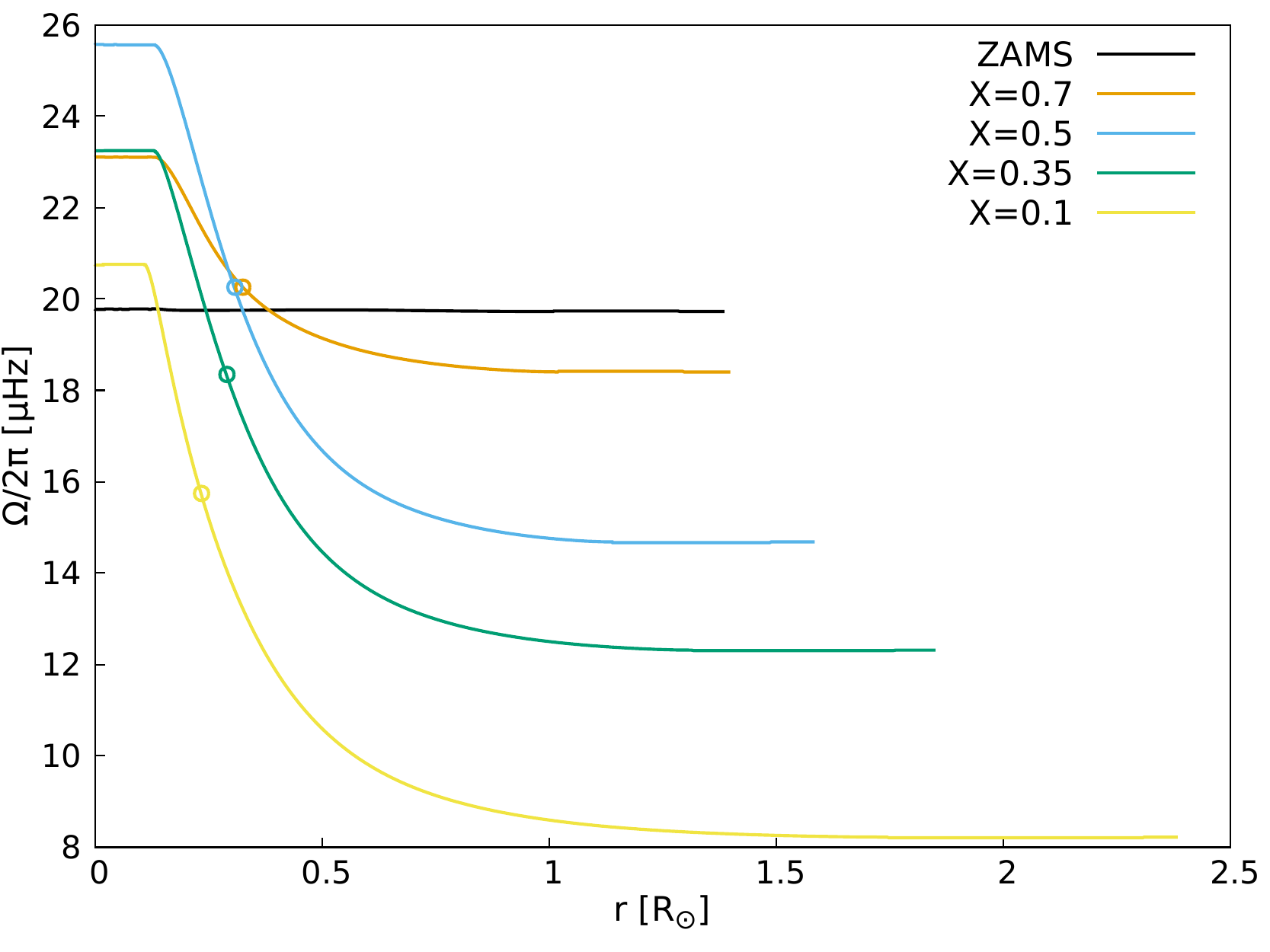}}
     \caption{Rotation rate as a function of the radial distance from the centre of the star in units of solar radii, for models without internal magnetic fields.
       The points over the lines indicate the mean rotation rate of the gravity-mode cavity computed with Eq. \ref{eq_meanomega} for each model.
       The models are shown at different stages during the MS, as indicated by the mass fraction of central hydrogen in the figure.}
     \label{rotprof_onlyrot}
   \end{figure}

     The evolution of the rotation rate in the stellar interior (i.e. the rotation profiles) through the MS without internal magnetic fields is shown in Fig. \ref{rotprof_onlyrot}.
     The central regions correspond to the convective core, so the rotation profile is flat there since we assume uniform rotation in convective regions, while the rotation rate in the outer regions decreases mainly as a consequence of the radial expansion through the MS.
     The rotation profile reaches a pseudo-equilibrium state early during the MS where the convective core rotates faster than the radiative interior as a result of the meridional circulation and the shear instability. Thus, its morphology is relatively unaffected until the end of the MS.
   %
   %
   \section{Internal magnetic fields allow for small inner differential rotation}
\begin{figure}[h!]
     \resizebox{\hsize}{!}{\includegraphics{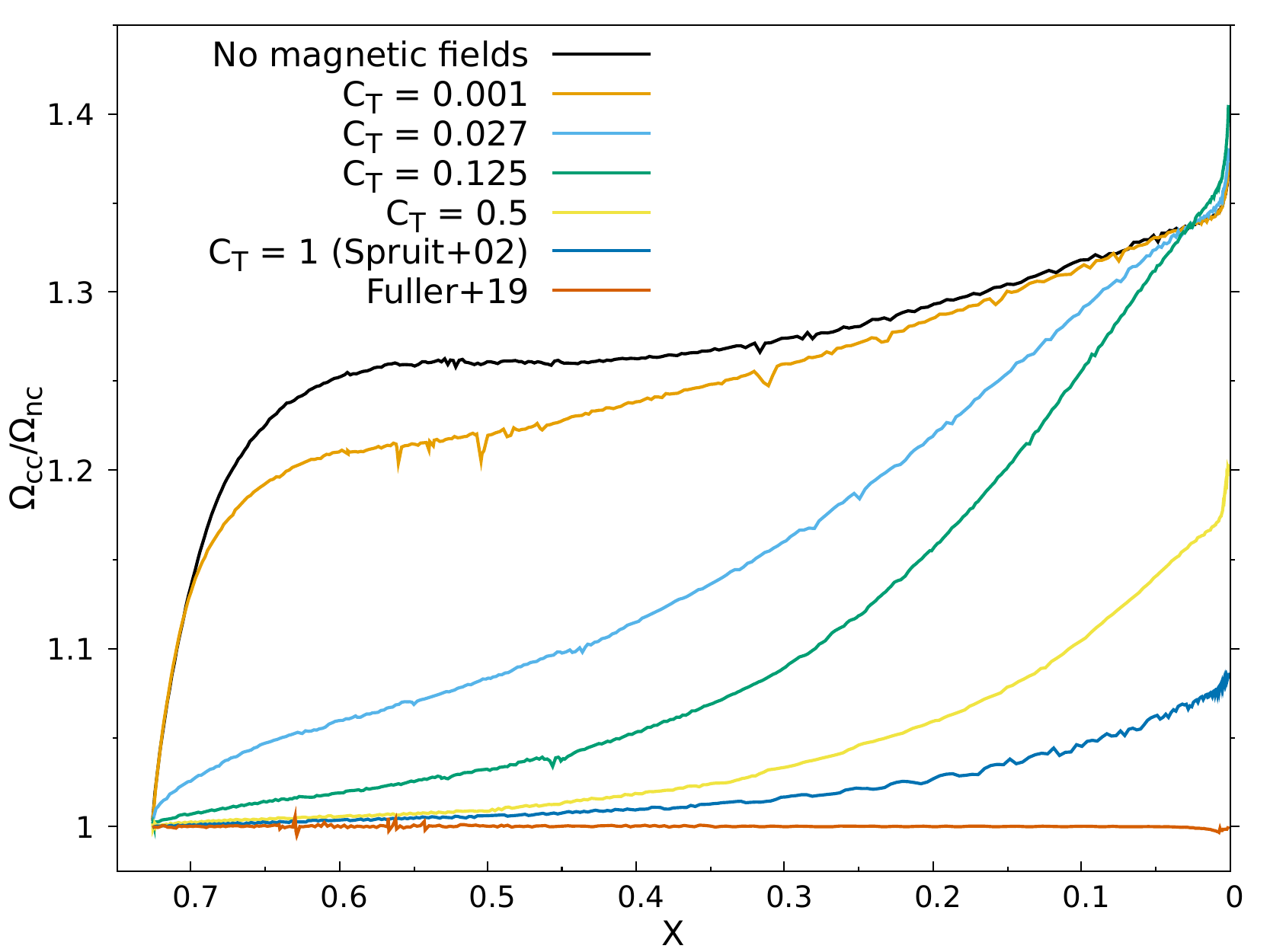}}
     \caption{Same as Fig. \ref{innerdr_xc_multiplot}, but for models including internal magnetic fields.
         The efficiency of angular momentum transport is proportional to $C_{\rm T}$ (See Sect. \ref{physics}).
     Models without magnetic fields and models computed with the original prescriptions proposed by \citet{spruit02} and \citet{fuller19} are also included for reference.
     }
     \label{innerdr_magnetic}
   \end{figure}

   Models with internal magnetic fields develop less inner differential rotation than models with purely hydrodynamical processes.
   In Fig. \ref{innerdr_magnetic} we show the evolution through the MS of the degree of inner differential rotation in models with internal magnetic fields and its sensitivity to the constant $C_{\rm T}$ (see Eqs. \ref{eq_qmin} and \ref{eq_numag}).
     Since the constant $C_{\rm T}$ sets the efficiency of the AM transport, low values of such a constant (e.g. $C_{\rm T}=0.001$) lead to similar evolution and degree of inner differential rotation as in models without magnetic fields, while high values lead to smaller inner differential rotation.
     Models with a weak AM transport efficiency converge to the same degree of inner differential rotation towards the end of the MS because the receding convective core leads to an extended region of relatively strong chemical gradients, which suppress the TS dynamo and so the convective core can decouple from the radiative interior as in models with only hydrodynamical processes.
     The model computed with $C_{\rm T}=1$ corresponds to the original prescription provided by \citet{spruit02}.
     Models computed with the prescription proposed by \citet{fuller19} lead to rigid rotation near convective-core boundaries.

\begin{figure}[h!]
     \resizebox{\hsize}{!}{\includegraphics{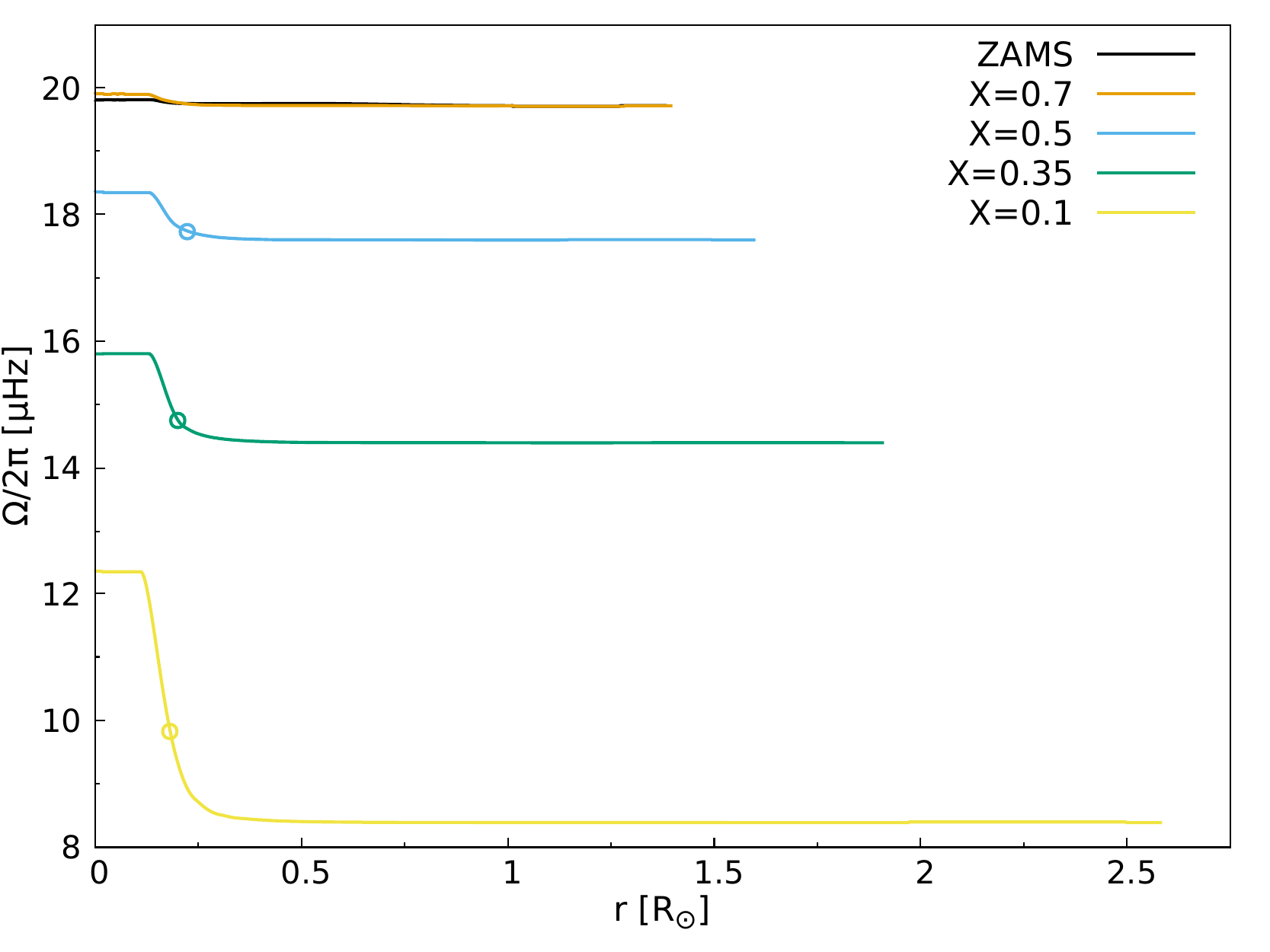}}
     \caption{Same as Fig. \ref{rotprof_onlyrot}, but for models including internal magnetic fields following a modified and less efficient version of the Tayler-Spruit dynamo, which corresponds to models computed with $C_{\rm T}=0.125$ shown in Fig. \ref{innerdr_magnetic}.}
     \label{rotprof_n1a05}
   \end{figure}

   In Fig. \ref{rotprof_n1a05}, we show an example of the kind of rotation profiles during the MS of our models with internal magnetic fields using a value of $C_{\rm T}=0.125$ in Eqs. \ref{eq_qmin} and \ref{eq_numag}.
   As the stars evolve through the MS, the convective core detaches from the radiative
   interior because the convective core retreats; thus, the gradients of chemical composition develop close to the convective boundaries.
   Furthermore, since the TS dynamo is inhibited by chemical composition gradients, thus differential rotation close to the boundaries of the convective core can develop progressively through the MS.
   In the outer radiative interior and up to the surface, the rotation profile is nearly uniform since in chemically homogeneous regions the dynamo can operate efficiently and even more far from the centre (see Eq. \ref{eq_numag}).  
   However, this kind of profile is only possible when adopting values of $C_{\rm T} \lesssim 0.5$.
   A value of $C_{\rm T}=216$, as needed to reproduce the core rotation rate of red giants \citep{eggenberger22}, leads to almost rigid rotation during the MS. The same conclusion is reached when using the prescription provided by \citet{fuller19}.
   %
   %
      \section{Comparison with asteroseismic constraints}
      We compare the degree of inner differential rotation and near-core rotation rates obtained in our models with the estimates provided by \citet{saio21} and \citet{li20}.
      We chose these quantities because they are not strongly sensitive to stellar parameters: the near-core rotation rate derived from the slope of the period spacing patterns by \citet{li20} is robust to changes in mass, metallicity, and evolutionary age \citep{ouazzani17}. In addition, the inner differential rotation as measured by the ratio $\Omega_{\rm cc}/\Omega_{\rm nc}$ is robust compared to changes in stellar parameters extracted from the best-fit models presented by \citet{saio21}, such as central hydrogen abundance or stellar mass.
      For the comparison with the models, we take the rotation rate of the convective core estimated from the best-fit models without overshooting presented by \citet{saio21} and the near-core rotation rates provided by \citet{li20}.
      The uncertainty on the rotation rate of the convective core is simply taken as the difference between the estimates obtained with models using a different overshooting strength by \citet{saio21}.
   \begin{figure}[hb!]
     \resizebox{\hsize}{!}{\includegraphics{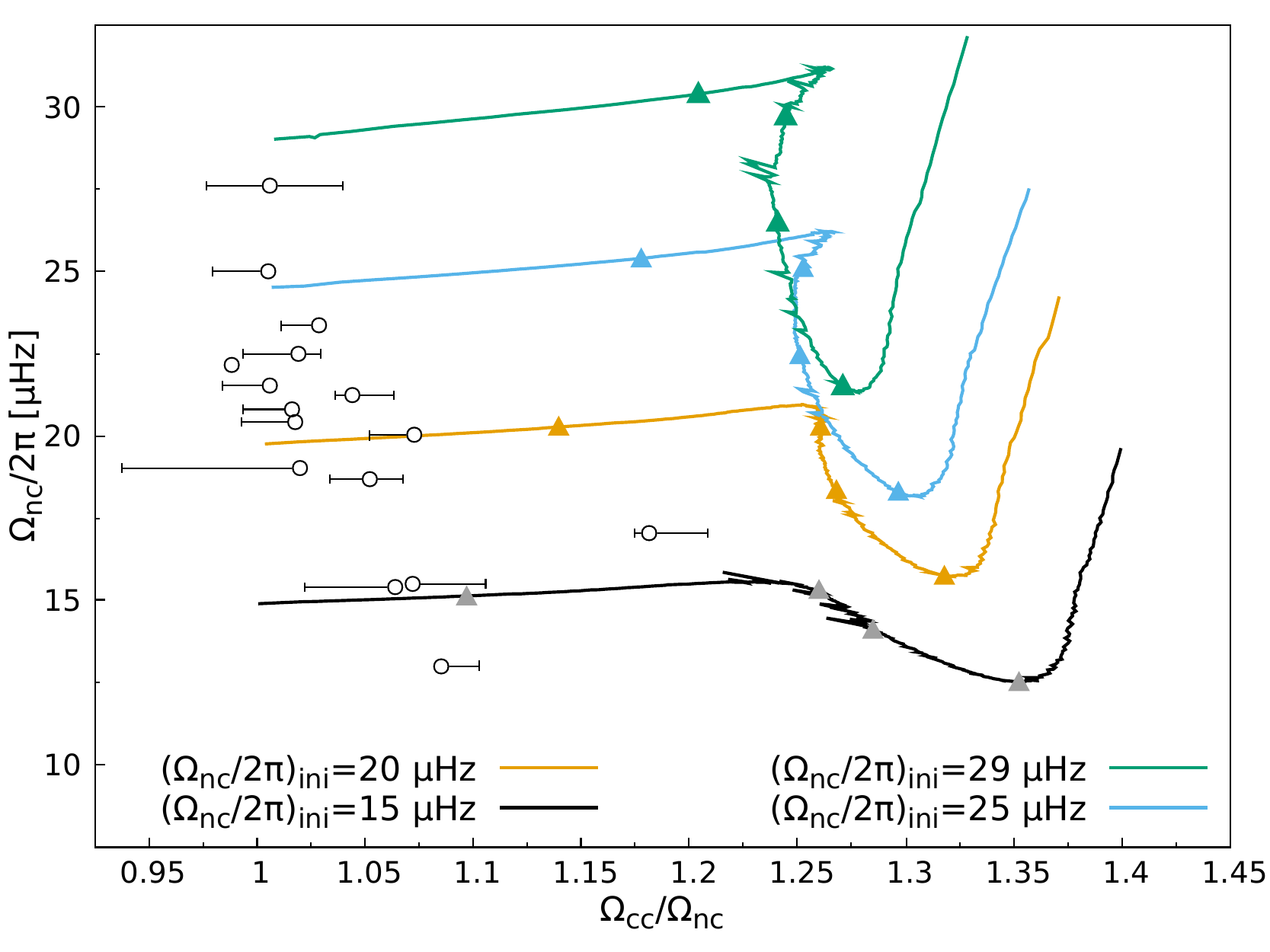}}
     \caption{Near-core rotation rate ($\Omega_{\rm nc}/2\pi$) as a function of the ratio of rotation rate between the convective core and the near-core radiative region ($\Omega_{\rm cc}/\Omega_{\rm nc}$).
       The models include angular momentum transport by meridional currents and shear instabilities, but do not include internal magnetic fields.
     The models are computed from the zero age main sequence assuming initial solid-body rotation $(\Omega_{\rm cc}/\Omega_{\rm nc})_{\rm ini} = 1$, with different initial rotation rates as indicated in the figure.
     The triangles over the evolutionary tracks indicate the time the central hydrogen content reaches $X=0.7, 0.5, 0.35,$ and $0.1$ (from left to right, respectively).
     The data points correspond to the constraints provided by \citet{saio21}.
}
     \label{omeganc_innerdr_onlyrot}
   \end{figure}

   In Fig. \ref{omeganc_innerdr_onlyrot}, we show that the degree of inner differential rotation in models without internal magnetic fields is in disagreement with the current constraints, since (as shown in Fig. \ref{rotprof_onlyrot}) the convective core decouples from the radiative interior early during the MS.
   This is illustrated by the points over the different curves located when the central abundance of hydrogen reaches $X=0.7, 0.5, 0.35,$ and 0.1.   
   For the model with an initial near-core rotation rate of $\Omega_{\rm nc}/2\pi = 29 \mu$Hz, as soon as the central hydrogen content reaches  $X \simeq 0.7,$ the convective core already rotates approximately 20\% faster than the radiative interior.
   For the lowest initial velocities, the points at $X=0.7$ are located progressively to the left because the convective core decouples slower at low initial velocities (as shown in Fig. \ref{innerdr_xc_multiplot}). This is because the efficiency of meridional circulation is proportional to the rotation rate.
   While, in principle, these models can potentially reproduce the constraints, we do not deem it very likely as the phase of readjustment of the rotation profile occurs relatively fast.

\begin{figure}[h!]
     \resizebox{\hsize}{!}{\includegraphics{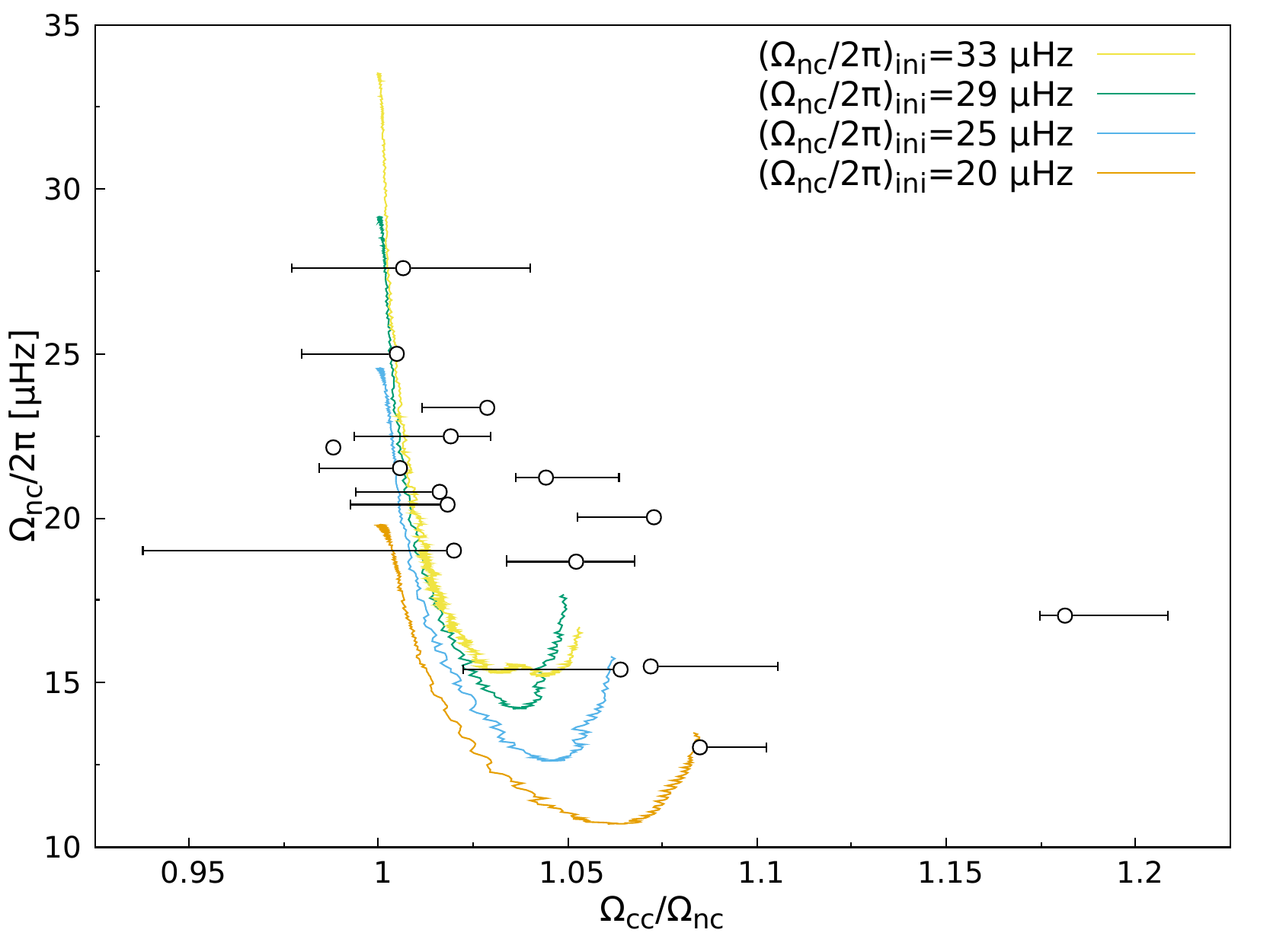}}
     \caption{Same as Fig. \ref{omeganc_innerdr_onlyrot}, but for models including internal magnetic fields following the original prescription of \citet{spruit02}.}
     \label{omeganc_innerdr_n1a1}
   \end{figure}

   In Fig. \ref{omeganc_innerdr_n1a1}, we show that models with internal magnetic fields following the original prescription by \citet{spruit02} cannot reproduce the degree of inner differential rotation and the near-core rotation rate simultaneously.
     These kinds of models cannot reproduce the observational constraints for any initial rotational velocity, due to the fact that the TS dynamo acts more efficiently at higher rotational velocities (see Eqs. \ref{eq_qmin} and \ref{eq_numag}).
     Because of this, the degrees of inner differential rotation displayed by the data cannot be reproduced by choosing high initial velocities for the models.
     In fact, our model with an initial rotation rate of $\Omega/2\pi = 33 \mu$Hz corresponds to $80$\% of its critical velocity and achieves a critical value at roughly the middle of the MS; this is also why the behaviour of this model is different towards the end of the MS, resulting from AM removal due to mechanical mass-loss.
   More efficient formulations, such as that proposed by \citet{fuller19}, lead to quasi-rigid rotation and, thus, they are not favoured by the data. As a result,  we find that a less efficient AM transport process is needed.

\begin{figure}[h!]
     \resizebox{\hsize}{!}{\includegraphics{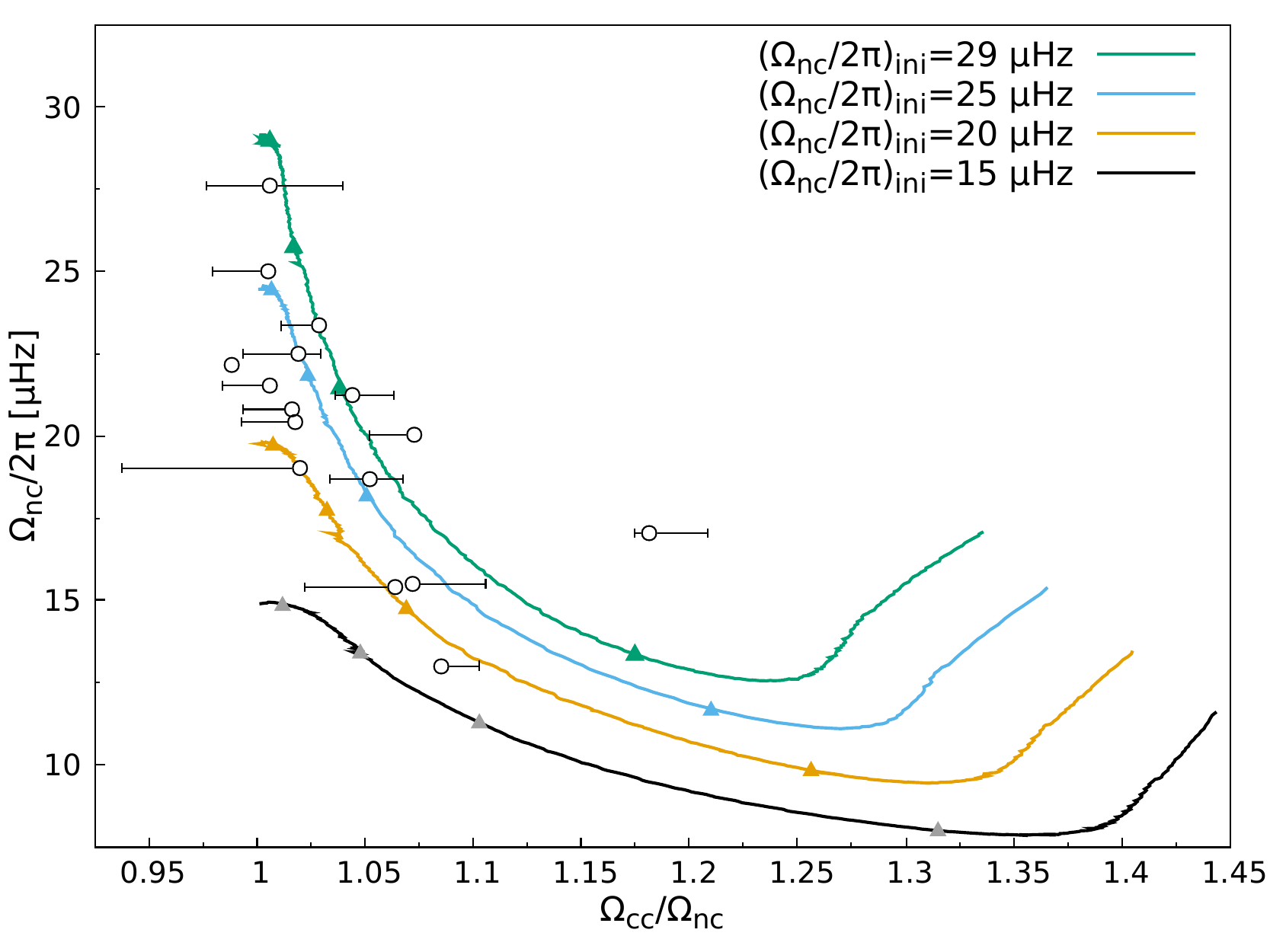}}
     \caption{Same as Fig. \ref{omeganc_innerdr_onlyrot}, but for models including internal magnetic fields following a modified and less efficient version of the Tayler-Spruit dynamo, computed with $C_{\rm T}=0.125$ in Eqs. \ref{eq_qmin} and \ref{eq_numag}.}
     \label{omeganc_innerdr_n1a05}
      \end{figure}

      In Fig. \ref{omeganc_innerdr_n1a05}, we show that our models with internal magnetic fields following a modified and less efficient version of the TS dynamo computed with $C_{\rm T}=0.125$ can reproduce the qualitative trends in a reasonable way.
      These models spend more than half of their MS lifetime with degrees of inner differential rotation in agreement with asteroseismic constraints, as illustrated by the symbols over the different tracks.
      This occurs because the convective core decouples from the radiative interior progressively due to the development of chemical gradients in the near-core regions and so, the inner differential rotation increases progressively as the convective core recedes.
      In the radiative regions above the core, the AM is redistributed efficiently, leading to a decreasing $\Omega_{\rm nc}/2\pi$. We note that the evolution of the rotation profile for this kind of model is shown in Fig. \ref{rotprof_n1a05}.
      The latter is needed to satisfy the constraints on the apparently quasi-rigid rotation from the surface to the near-core regions \citep{vanreeth18,li20} and to reproduce their surface rotational velocities \citep{moyano23b}.      
      Moreover, this conclusion is largely independent of the rotational history during the pre-MS, since before the hydrogen starts burning in the core, little to no chemical composition gradients are present in the radiative regions.
      This allows for the TS dynamo to operate efficiently, since buoyancy forces only work to weakly stabilise the radial motions associated to the Tayler instability when we are dealing with a chemically homogeneous medium.
      Therefore, our models with internal magnetic fields always converge to a uniform rotation profile by the time they arrive at the ZAMS.

      We note that our interpretation of the data is different from that of \citet{saio21}, who suggested that the convective core re-couples with the radiative interior through evolution; so, the degree of inner differential rotation decreases progressively, instead of increasing as it is shown to do in our models.
      However, that would imply that the near-core regions should rotate faster as the stars evolve through the MS, which we do not support since our previous studies have demonstrated that to reproduce the surface rotational velocities of these kinds of stars, the AM from the near-core regions should be transported to the surface \citep{moyano23b}.
      Moreover, evidence from previous studies using gravity modes and the modulation seen in the light curves (potentially caused by magnetic spots) suggest quasi-rigid rotation in radiative regions \citep{vanreeth18,li20}.
      This should lead to decreasing near-core rotation rates, as the AM is extracted from the near-core regions and redistributed in the whole radiative interior.

      %
      %
   \section{Implications for the angular momentum transport problem}
   \label{implications}
   Internal magnetic fields generated by the TS dynamo either in its standard  \citep{spruit02}, revised \citep{fuller19}, or calibrated forms \citep{eggenberger22} lead to quasi-rigid rotation and  are  rejected as a result.
    Purely solid-body rotation or our models with purely hydrodynamical processes  are not able to reproduce the qualitative trends seen in the data.
   Our models with a modified and less efficient version of the TS dynamo are in better agreement with the rotational constraints on the convective core and the radiative interior.
   However, this less efficient version leads to weaker transport during the post-MS phase, which, in turn, leads to core rotation rates that are in disagreement with asteroseismic constraints \citep[e.g. those from][]{mosser12,gehan18}.
   Moreover, \citet{deheuvels20} found that two young subgiants  just beginning to expand towards the red giant phase, rotate roughly as solid bodies. We find that our models cannot account for these constraints.
   This raises the question of whether the same physical process should dominate the AM transport during the MS and the subgiant or red giant phase.
   It is possible that other processes are at work during the post-MS phase efficient enough to slow down the cores.
   Some of these processes might include internal gravity waves \citep[e.g.][]{pincon17}, mixed modes \citep{belkacem15a,belkacem15b} that remain as-yet unexplored in stellar evolution calculations, or another form of magnetic fields or magnetic instabilities, such as the azimuthal magneto-rotational instability \citep[e.g.][]{meduri23} or large-scale internal magnetic fields \citep[e.g.][]{takahashi21}.
         
   
   \begin{acknowledgements}
         FDM thanks Masao Takata for useful discussions during the MIAPbP workshop held during August of 2023.
    The authors have received funding from the European Research Council (ERC) under the European Union’s Horizon 2020 research and innovation programme (grant agreement No. 833925, project STAREX).
    The computations were performed at University of Geneva using Baobab HPC service.
\end{acknowledgements}

\bibliographystyle{aa}
\bibliography{cc_gammadors}
\end{document}